\documentclass[notitlepage,
superscriptaddress,showpacs,
twocolumn]{revtex4-1}
\usepackage{hyperref}
\usepackage{graphicx}
\usepackage{mathrsfs}
\usepackage{amsmath}

\begin{document}
\title{Quantum correlations in
critical $XXZ$ system and LMG model}

\author{Biao-Liang Ye}
\affiliation{School of Physics and
Electronic Information, Shangrao Normal
University, Shangrao 334001, China}
\author{Bo Li}
\affiliation{School of Mathematics and
Computer Sciences, Shangrao Normal
University, Shangrao 334001, China}
\affiliation{Max-Planck-Institute for
Mathematics in the Sciences, Leipzig 04103, Germany}
\author{Xianqing Li-Jost}
\affiliation{Max-Planck-Institute for
Mathematics in the Sciences, Leipzig 04103, Germany}
\author{Shao-Ming Fei}
\affiliation{Max-Planck-Institute for
Mathematics in the Sciences, Leipzig 04103, Germany}
\affiliation{School of Mathematical Sciences,
Capital Normal University, Beijing 100048,
China}

\begin{abstract}
We investigate the quantum phase transitions for the $XXZ$ spin-1/2
chains via the quantum correlations between the nearest and next to nearest neighbor spins characterized by
negativity, information deficit, trace distance discord and
local quantum uncertainty. It is shown that all these correlations
exhibit the quantum phase transitions
at $\Delta=-1$. However, only
information deficit and local
quantum uncertainty can demonstrate
quantum phase transitions at
$\Delta=1$. The analytical
and numerical behaviors of the
quantum correlations for the
$XXZ$ system are presented. We
also consider quantum correlations in
the Hartree-Fock ground state of
the Lipkin-Meshkov-Glick (LMG) model.
\end{abstract}
\date{\today}
\maketitle

\section{Introduction}
Quantum entanglement is an ubiquitous
resource in quantum information processing \cite{Amico2008} and has
many significant applications in quantum information
tasks \cite{Horodecki2009}.
Besides quantum entanglement, there are also quantum correlations
that can be used to realize some
quantum speed up without entanglement \cite{Merali2011}.
Much attentions have been paid to all such nonclassical
correlations \cite{Modi2012}. Many measures to quantify nonclassical correlations
have been proposed \cite{Adesso2016}, including
quantum discord \cite{Ollivier2001} and
information deficit \cite{Oppenheim2002}.
A geometric method in quantifying quantum discord
has been provided in \cite{Dakic2010}.
The analytical expressions of quantum
correlation for arbitrary two-qubit $X$ states
have been presented via trace distance discord \cite{Ciccarello2014}.
Inspired by Wigner-Yanase Skew information,
in Ref.\cite{Girolami2013} the local quantum uncertainty is proposed
to quantify nonclassical correlations.

On the other hand, quantum phase transition is a fundamental
phenomena in condensed matter physics, and is tightly
related to quantum correlations.
In Ref. \cite{Werlang2010,Werlang2011},
the authors utilize the entanglement of formation and quantum discord
to spotlight the quantum critical points
for the $XXZ$ model, $XY$ model, and the Ising model with
external magnetic field at finite temperatures.
In Ref. \cite{Li2011} the authors used quantum discord and classical correlation
to detect quantum phase transitions for $XY$
spin chain with three-spin interactions at both
zero and finite temperatures.
The authors in \cite{Justino2012} revealed a general quantum phase transition
in an infinite one-dimensional $XXZ$ chain in terms of
concurrence and Bell inequalities.
In terms of the density matrix renormalization
group theory method, the behaviors of the
quantum discord, quantum coherence and Wigner-Yansase
skew information the relations between the phase transitions and symmetry points
in the Heisenberg $XXZ$ spin-1 chains have been extensively investigated in \cite{Malvezzi2016}.
The classical correlation and quantum discord also
exhibit the signatures of the quantum phase transitions
of $XXZ$ model and the Lipkin-Meshkov-Glick (LMG) model \cite{Sarandy2009}.

Although the Heisenberg $XXZ$ model has
been extensively investigated from different perspectives owning to
its rich physics, the quantum correlations used in studying quantum phase transitions
concern almost only concurrence and quantum discord.
It would be interesting if other quantum correlations
can also reveal general quantum phase transitions.
In this work, we take negativity, information deficit, trace distance
discord, and local quantum uncertainty to study the quantum phase transitions of the $XXZ$ spin-1/2 chains, as well as
the quantum correlations in the Hartree-Fock ground state of the Lipkin-Meshkov-Glick model.
In sect. II, we review several basic notations and concepts
of measures of quantum correlations.
In sect. III, the Heisenberg $XXZ$ model is introduced.
We discuss the computation of the quantum correlations and
illustrate the results for the XXZ chain.
In sect. IV, analytical and numerical results for all the quantum
correlations are demonstrated for the LMG model.
Finally, we conclude in sect. V.

\section{Measures of quantum correlations}
Let us first review the basic notations
and concepts of several quantum correlation measures.

{\sf Negativity}~
Negativity is a computable measure of quantum
entanglement \cite{Vidal2002}. It can
be calculated effectively for any mixed
states of arbitrary bipartite systems.
The negativity $\mathcal{N}(\varrho_{AB})$ of a bipartite state $\varrho_{AB}$ is defined by
\begin{equation}\label{N}
\mathcal{N}=\frac{\|\varrho^{T_A}\|_1-1}{2}=|\sum_i\mu_i|,
\end{equation}
where $\varrho^{T_A}$ is the partially
transposed state of $\varrho_{AB}$ with respect to the subsystem $A$, $\|\sigma\|_1=\textrm{Tr}\sqrt{\sigma^\dagger\sigma}$
denotes the trace norm one (or Schatten one-norm) of an Hermitian
operator $\sigma$, and $\mu_i$ are the negative eigenvalues of
$\varrho^{T_A}$ show how much $\varrho^{T_A}$ fails to be positive definite.

{\sf Information deficit}
Let $\Pi_j^A$ denote a local complete projective measurement
on subsystem $A$, which satisfy
$\sum_j\Pi_j^A=I_A$ and $\Pi_j^A\Pi_k^A=\delta_{jk}\Pi_k^A$, with $I_A$ being the
identity operator on subsystem $A$.
For the case that $\varrho_A=\textrm{Tr}_B(\varrho_{AB})$ is
a single-qubit state, the rank-1 projectors
are of the form
$\Pi_j^A=|\Omega_j\rangle\langle \Omega_j|$,
$j=0,1,$ where
\begin{eqnarray}\label{bases}
 	|\Omega_0\rangle&=&\cos\theta|0\rangle
 	+e^{i\varphi}\sin\theta|1\rangle,\nonumber\\
 	|\Omega_1\rangle&=&-e^{-i\varphi}\sin
 	\theta|0\rangle+\cos\theta|1\rangle,
 \end{eqnarray}
with $0\le\theta\le\pi$, $0\le\varphi<
2\pi$, and $\{|0\rangle, |1\rangle\}$ the computational basis of the subsystem $A$.

After the projective measurement, the state $\varrho_{AB}$ of the total system becomes
\begin{equation}\label{ps}
	\varrho_{AB}'=\sum_j\Pi_j^A\varrho_{AB}
	\Pi_j^A.
\end{equation}
The information deficit
is the minimum information loss by the measurement,
\begin{equation}\label{I}
	\mathcal{I}(\varrho_{AB})=\min_{\Pi_j^A}S(\varrho_
	{AB}')-S(\varrho_{AB}),
\end{equation}
where $S(\varrho)=-\mathrm{Tr}\varrho
\log\varrho$ is the von Neumann entropy and
$\log$ is in base 2.
Here we use the
definition of information deficit instead
of the one-way information deficit
\cite{Oppenheim2002,Horodecki2005,Streltsov2011}
for brevity. Similar to the
quantum discord \cite{Ollivier2001},
the analytical expressions for the
information deficit of the simplest
two-qubit states are still not known yet \cite{Ye2016}.

{\sf Trace distance discord}~
The definition of trace distance discord is given by
\begin{equation}\label{D}
	\mathcal{D}(\varrho_{AB})=\frac12\min_{\Pi_j^A}\|\varrho_{AB}
	-\varrho_{AB}'\|_1.
\end{equation}
Here, the trace norm one $\|\cdot\|_1$ is the same as the one in Eq.(\ref{N}), and the $\varrho_{AB}'$
is the state after measurement on subsystem $A$ in Eq.(\ref{ps}).

Trace distance discord is a reliable
geometric quantifier of discord-like
correlations \cite{Ciccarello2014}. It provides an explicit
and compact expression for two-qubit $X$ states.

{\sf Local quantum uncertainty}
Uncertainty of local observables for a
bipartite system is a \emph{bona fide} measure
of nonclassical correlation. Local
quantum uncertainty can play important
roles in the context of quantum metrology.

Local quantum uncertainty is the minimum
skew information \cite{Wigner1963,Luo2003} achievable by local
measurements. The minimum achievable skew
information by a single local measurement is given by
\begin{equation}
\mathcal{U}=\min_{K^\Gamma}I(\varrho, K^\Gamma),
\end{equation}
where $\Gamma$ denotes the spectrum of $K^\Gamma$,
and the minimization over a chosen spectrum of
observables leads to a specific measure from
the family and $I(\varrho, K^\Gamma)=-\frac12{\rm Tr}([\sqrt{\varrho}, K^\Gamma]^2)$.
However, for a two-qubit system, all
of the members of the family turn out
to be equivalent. For two-qubit system, the
local quantum uncertainty admits a closed
formula.
The local quantum uncertainty with respect
to subsystem $A$ can be derived by
\begin{equation}\label{U}
	\mathcal{U}=1-\lambda_{\max}\{W_{AB}\},
\end{equation}
where $\lambda_{\max}$ is the maximum
eigenvalues and $W_{AB}$ denotes a
$3\times3$ symmetric matrix whose elements
are given by
\begin{equation}
	(W_{AB})_{uv}=\mathrm{Tr}
	\{\varrho_{AB}^{1/2}(\sigma_{A}^u\otimes I_B)\varrho_{AB}^{1/2}(\sigma_{A}^v\otimes I_B)\},
\end{equation}
with $\sigma_{A}^{u(v)}$ being the Pauli matrices, $u,v=x,y,z.$

\section{Heisenberg $XXZ$ spin-1/2 chain}


We now consider the one-dimensional spin
chain with anisotropic Heisenberg interactions.
The Hamiltonian of the $XXZ$ model is given by
\begin{equation}\label{ha}
	H=\sum_{j=1}^N[S_j^xS_{j+1}^x+S_j^yS_{j+1}^y+\Delta S_j^zS_{j+1}^z],
\end{equation}
where $S_j^u=\sigma_j^u/2$ $(u=x,y,z),$
$\sigma_j^u$ are the Pauli operators
on site $j$, $\Delta$ is the anisotropic
parameter, $\sigma_{j+N}^u=\sigma_j^u$,
and $N$ is the number of spins of
the chain. For $T=0$ the $XXZ$ model has two
critical points \cite{Takahashi2005}.
The first-order
transition happens at $\Delta=-1$ and a
continuous phase transition shows up at $\Delta=1$.

Due to symmetry in the spin chain model with Hamiltonian Eq.(\ref{ha}),
the two-qubit reduced
density matrix of sites $i$ and $i+r$ in the
basis $|1\rangle=|\uparrow\uparrow\rangle$,
$|2\rangle=|\uparrow\downarrow\rangle$,
$|3\rangle=|\downarrow\uparrow\rangle$,
$|4\rangle=|\downarrow\downarrow\rangle$
(where $|\uparrow\rangle$ and $|\downarrow\rangle$
are the eigenstates of the Pauli spin $z$-operator)
has the following form,
\begin{eqnarray}
	\varrho_{AB}=\left(
	\begin{array}{cccc}
		\varrho_{11} & 0 & 0 & 0\cr
		0 & \varrho_{22} & \varrho_{23} & 0\cr
		0 & \varrho_{32} & \varrho_{33} & 0\cr
		0 & 0 & 0 & \varrho_{44}
	\end{array}
	\right),
\end{eqnarray}
with
$$
	\varrho_{23}=\varrho_{32}=\frac{\langle\sigma_i^x\sigma_{i+r}^x\rangle}{2},
$$
$$
	\varrho_{11}=\varrho_{44}=\frac{1+\langle\sigma_i^z\sigma_{i+r}^z\rangle}{4},
$$
and
$$
	\varrho_{22}=\varrho_{33}=\frac{1-\langle\sigma_i^z\sigma_{i+r}^z\rangle}{4}.
$$

\emph{The correlation functions for nearest neighbor (r=1) spins of $XXZ$ model}.
The two point correlation functions of the $XXZ$ model at zero temperature and in the thermodynamics limit
can be derived by using the Bethe ansatz technique \cite{Shiroishi2005}.
The spin-spin correlation functions between nearest-neighbor
spin sites for $\Delta>1$ are given by Takahashi et al. \cite{Takahashi2004}
\begin{widetext}
\begin{eqnarray}
	\langle\sigma_i^z\sigma_{i+1}^z\rangle&=&
	1+2\int_{-\infty+i/2}^{\infty+i/2}
	\frac{dx}{\sinh(\pi x)}(\cot(\nu x)\coth(v)
	-\frac{x}{\sin^2(\nu x)}),
\end{eqnarray}
and
\begin{eqnarray}
	\langle\sigma_i^x\sigma_{i+1}^x\rangle&=&
	\int_{-\infty+i/2}^{\infty+i/2}\frac{dx}{\sinh(\pi x)}(\frac{x}{\sin^2(\nu x)}\cosh\nu-\frac{\cot(\nu x)}{\sinh\nu}),
\end{eqnarray}
\end{widetext}
with $\nu=\cosh^{-1}\Delta.$
For $\Delta=1$, $\langle\sigma_i^x\sigma_{i+1}^x\rangle=\langle\sigma_i^z\sigma_{i+1}^z\rangle=1/3(1-4\ln 2)$, and for
$\Delta\le -1$, $\langle\sigma_i^z\sigma_{i+1}^z\rangle=1$
and $\langle\sigma_i^x\sigma_{i+1}^x\rangle=0$ \cite{Justino2012}.

For $-1<\Delta<1$, the correlation
functions are given by Kato et al. \cite{Kato2004}
\begin{widetext}
\begin{eqnarray}
	\langle\sigma_i^z\sigma_{i+1}^z\rangle &=& 1 -
	\frac{2}{\pi^2}\int_{-\infty}^\infty\frac{dx}{\sinh x}\frac{x \cosh x}{\cosh^2(\Phi x)}
	+\frac{2\cot(\pi\Phi)}{\pi}\int_{-\infty}^\infty\frac{dx}{\sinh x}\frac{\sinh((1-\Phi)x)}{\cosh(\Phi x)},\nonumber\\
\end{eqnarray}
and
\begin{eqnarray}
	\langle\sigma_i^x\sigma_{i+1}^x\rangle &=& \frac{\cos(\pi\Phi)}{\pi^2}\int_{-\infty}^{\infty}
	\frac{dx}{\sinh x}\frac{x \cosh x}{\cosh^2(\Phi x)}- \frac{1}{\pi\sin(\pi\Phi)}\int_{-\infty}
	^\infty\frac{dx}{\sinh x}\frac{\sinh((1-\Phi)x)}{\cosh(\Phi x)},
\end{eqnarray}
\end{widetext}
with $\Phi=\frac{1}{\pi}\cos^{-1}\Delta.$

\begin{figure*}[htbp!]
\includegraphics[width=7.2in]{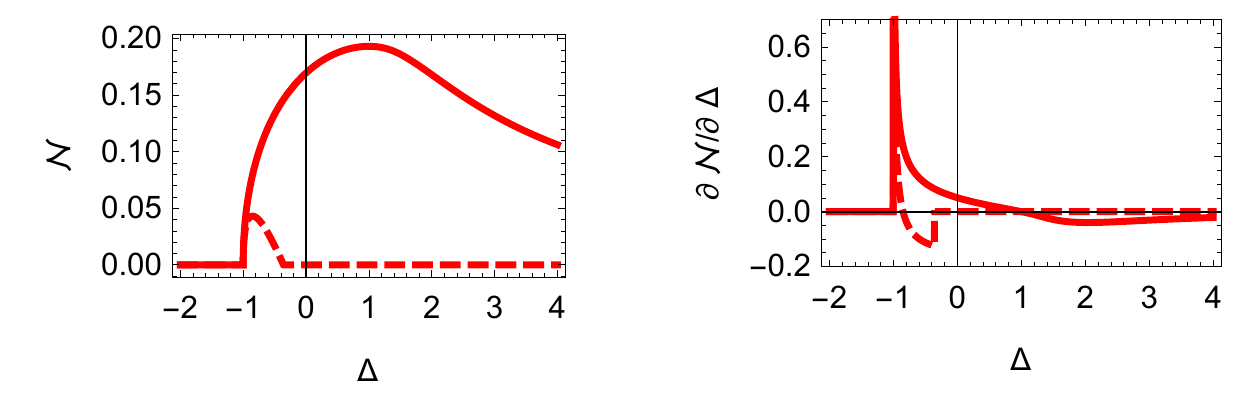}\\
\centering
\caption{(Color online) Negativity (left)
and its derivative with respect to $\Delta$ (right) vs $\Delta$ for nearest neighbor spins (solid line) and
next to nearest neighbor spins (dashed line) of the $XXZ$ model.}
\label{fig1}
\end{figure*}

\emph{The correlation functions for next to nearest neighbor (r=2) spins of $XXZ$ model}.
For the next to nearest-neighbor spins,
in the region $\Delta>1$, we have the correlation
functions \cite{Takahashi2004},
\begin{widetext}
\begin{eqnarray}
	\langle \sigma_i^x \sigma_{i+2}^x\rangle
	&=& \int_{-\infty+i/2}^{\infty+i/2}
	\frac{dx}{\sinh(\pi x)}\frac12[-\frac{x}{\sin^2(\nu x)}(\frac{3\sinh^2\nu}{\sin^2(\nu x)}
	+1-3\cosh2\nu)+\cot(\nu x)(\frac{3\cosh(2\nu)\tanh(\nu)}{\sin^2(\nu x)}-\frac{4}{\sinh(2\nu)})],\nonumber
\end{eqnarray}
and
\begin{eqnarray}
	\langle \sigma_i^z\sigma_{i+2}^z\rangle
	&=& 1+\int_{-\infty+i/2}^{\infty+i/2}
	\frac{dx}{\sinh(\pi x)}[\frac{x}{\sin^2(\nu x)}
	(\frac{3\sinh^2\nu}{\sin^2(\nu x)}-1-\cosh (2\nu))
	-\cot(\nu x)(\frac{3\tanh\nu}{\sin^2(\nu x)}
	-4\coth(2\nu))].
\end{eqnarray}
\end{widetext}
with $\nu=\cosh^{-1}\Delta.$
For
$\Delta\le -1$, $\langle\sigma_i^z\sigma_{i+1}^z\rangle=1$
and $\langle\sigma_i^x\sigma_{i+1}^x\rangle=0$.
If $\Delta=1$ \cite{Shiroishi2005}, then
$
\langle\sigma_i^x\sigma_{i+2}^x\rangle=\langle \sigma_i^z\sigma_{i+2}^z\rangle
=0.242719.
$
For $-1<\Delta< 1$ one has \cite{Kato2003}
\begin{widetext}
\begin{eqnarray}
	\langle\sigma_i^x\sigma_{i+2}^x\rangle
	&=&-\int_{-\infty}^\infty\frac{dx}{\sinh x}
	\frac{\sinh(1-\Phi)x}{\cosh(\Phi x)}[\frac{2}{\pi\sin(2\pi\Phi)}+\frac{3\cos2\pi\Phi\tan\pi\Phi}{\pi^3}x^2]\nonumber\\
	&+&\int_{-\infty}^{\infty}\frac{dx}{\sinh x}\frac{\cosh x}{(\cosh\Phi x)^2}[\frac{\cos2\pi\Phi}{\pi^2}x+\frac{(\sin\pi\Phi)^2}{\pi^4}x^3]
\end{eqnarray}
and
\begin{eqnarray}
	\langle\sigma_i^z\sigma_{i+2}^z\rangle
	&=&1+4\int_{-\infty}^\infty\frac{dx}{\sinh x}
	\frac{\sinh(1-\Phi)x}{\cosh\Phi x}
	[\frac{\cot(2\pi\Phi)}{\pi}+\frac{3\tan\pi\Phi}{2\pi^3}x^2]\nonumber\\
	&-&4\int_{-\infty}^\infty\frac{dx}{\sinh x}
	\frac{\cosh x}{(\cosh\Phi x)^2}[\frac{x}{2\pi^2}
	+\frac{(\sin\pi\Phi)^2}{2\pi^4}x^3].
\end{eqnarray}
\end{widetext}

i) By the definition of \emph{negativity}
Eq.(\ref{N}),  we have for the nearest neighbor spins of $XXZ$ model,
\begin{equation}
	\mathcal{N}=-\frac{1}{4} (1+\langle\sigma_i^z\sigma_{i+1}^z\rangle+2 \langle\sigma_i^x\sigma_{i+1}^x\rangle).
\end{equation}
For the next to nearest neighbor spins,
the analytical formula of negativity about the
$XXZ$ model is given by
\begin{equation}
	\mathcal{N}=-\frac{1}{4} (1+\langle\sigma_i^z\sigma_{i+2}^z\rangle-2 \langle\sigma_i^x\sigma_{i+2}^x\rangle),
\end{equation}
in region $\Delta\in(-1,-0.358733)$, and $\mathcal{N}=0$ for others region of $\Delta$, see
Fig. (\ref{fig1}) for the behaviors of quantum entanglement $\mathcal{N}$.

\begin{figure*}[htbp!]
\includegraphics[width=7.2in]{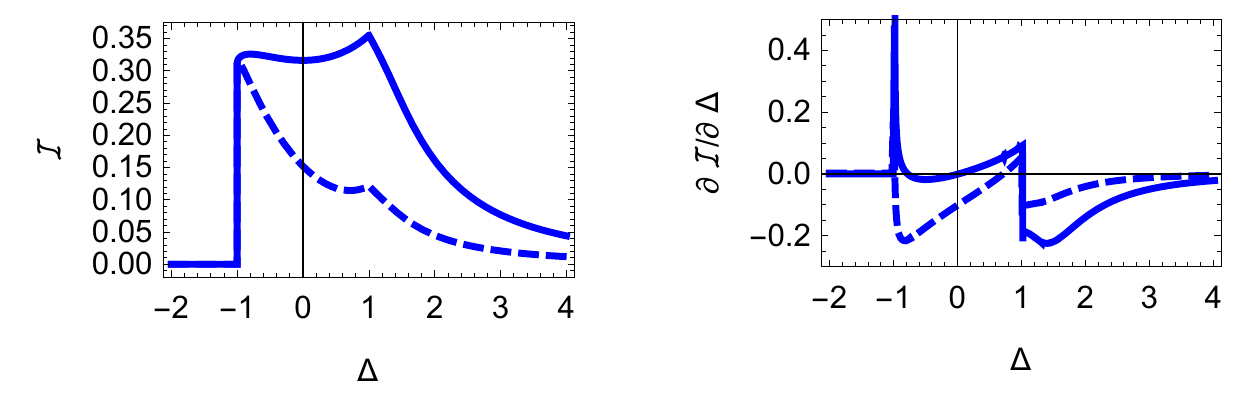}\\
\centering
\caption{(Color online) Information deficit (left) for the
$XXZ$ model for both nearest (solid line) and next to nearest (dashed line) neighbor spins and its derivative with respect to $\Delta$ (right) vs $\Delta$.}
\label{fig2}
\end{figure*}

From Fig. (\ref{fig1}), we observe that, for the
nearest neighbor spins, the negativity increases
monotonously with anisotropy in the region
$-1<\Delta<1$, while
in the region $\Delta>1$ the negativity decreases
as anisotropy increases. And the entanglement
reaches the maximal value tt $\Delta=1$.
For the next to nearest neighbor spins,
only in the region $\Delta\in(-1,-0.358733)$ the entanglement is non-zero.
Moreover, the sudden birth of entanglement happens at $\Delta=-1$ both for the
nearest and next to nearest neighbor spins.
From the right figure one can see that the transition happens also at the critical point $\Delta=-1$.

ii) We take the \emph{information deficit} to capture the quantum
correlation of the $XXZ$ system. In the parameter region $-1<\Delta<1$,
the optimal projective measurement bases (\ref{bases}) are obtained at $\theta=\pi/4$ and $\varphi=0$. We have
the analytical expression of information deficit for the nearest neighbor spins,
\begin{widetext}
\begin{eqnarray}
	\mathcal{I}&=&\frac14\{2(1+ \langle\sigma_i^z\sigma_{i+1}^z\rangle) \log (1+\langle\sigma_i^z\sigma_{i+1}^z\rangle)-
	\sum_\pm[2 (1\pm\langle\sigma_i^x\sigma_{i+1}^x\rangle) \log (1\pm\langle\sigma_i^x\sigma_{i+1}^x\rangle\nonumber)\\[2mm]
	&-&(1-\langle\sigma_i^z\sigma_{i+1}^z\rangle\pm2 \langle\sigma_i^x\sigma_{i+1}^x\rangle) \log (1-\langle\sigma_i^z\sigma_{i+1}^z\rangle\pm2 \langle\sigma_i^x\sigma_{i+1}^x\rangle)]\}.
\end{eqnarray}
For other regions of $\Delta$, the optimal measurement are given by $\theta=0$ and $\varphi=0$.
We have
\begin{eqnarray}
	 \mathcal{I}&=&\frac14\{
	\sum_\pm[(1-\langle\sigma_i^z\sigma_{i+1}^z\rangle\pm2 \langle\sigma_i^x\sigma_{i+1}^x\rangle) \log (1-\langle\sigma_i^z\sigma_{i+1}^z\rangle\pm2 \langle\sigma_i^x\sigma_{i+1}^x\rangle)\nonumber\\[2mm]
	&-&2(1- \langle\sigma_i^z\sigma_{i+1}^z\rangle) \log (1-\langle\sigma_i^z\sigma_{i+1}^z\rangle)]\}.
\end{eqnarray}
\end{widetext}
The analytical expression of information deficit for the next to nearest neighbor spins is in coincidence with the nearest neighbor spins. 
However, their scales are different (see Fig.\ref{fig2}). In fact, the above exact expressions of information deficit are also adapted for quantum discord \cite{Dillenschneider2008}.

From Fig.(\ref{fig2}), we see that the
information deficit in the region $\Delta>1$
decreases monotonously with anisotropy $\Delta$ for both
the nearest and next to nearest neighbor spins.
There are two critical points at $\Delta=-1$ and $1$ in the derivative
of information deficit with respect to $\Delta$, showing a kind of quantum phase transitions
for the $XXZ$ spin-1/2 chain.

iii) \emph{Trace distance discord}.
For the nearest neighbor spins, the analytical expression of trace distance discord is given by
$\mathcal{D}=|\langle\sigma_i^x\sigma_{i+1}^x\rangle|/2$,
which is also the formula for the next to nearest neighbor spins.

\begin{figure*}[htbp!]
\includegraphics[width=7.2in]{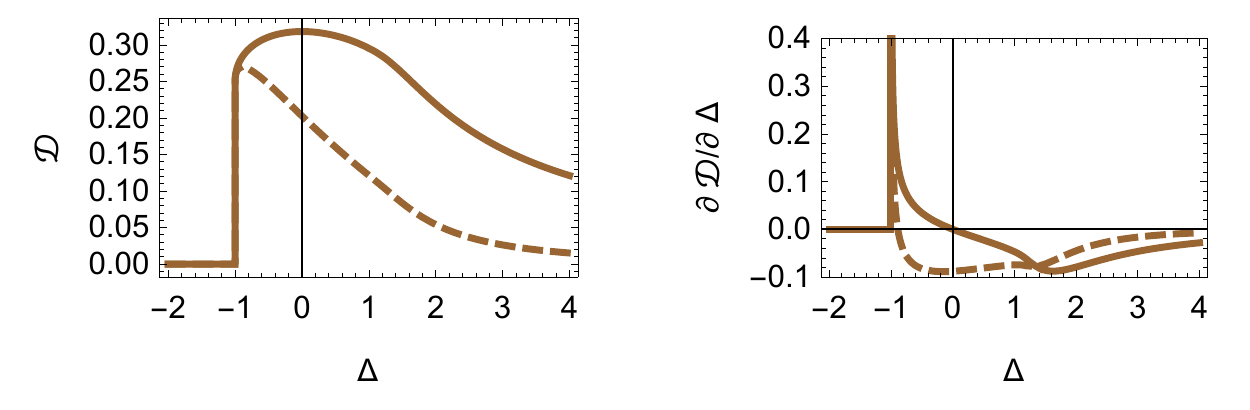}\\
\centering
\caption{(Color online) Trace distance discord (left)
and its derivative with respect to $\Delta$ (right) vs $\Delta$ for the
$XXZ$ model for nearest (solid line) and next to nearest (dashed line) neighbor spin.}
\label{fig3}
\end{figure*}

For the nearest neighbor spins,
the trace distance discord increases
monotonously with anisotropy in the region
$-1<\Delta<0$, while
in the region $\Delta\ge0$, it decreases
as anisotropy $\Delta$ increases (see Fig. (\ref{fig3})).
A sudden change happens at $\Delta=-1$ both for the
nearest and next to nearest neighbor spins, giving rise to quantum phase transitions of the $XXZ$ model.

iv) The analytical formula of the \emph{local quantum
uncertainty} for both nearest and next to nearest neighbor
spins can be written as
\begin{widetext}
\begin{eqnarray}
	\mathcal{U}=1-\left\{\begin{array}{ll}
	\sqrt{1+\langle\sigma_i^z\sigma_{i+1}^z\rangle}
	\left(\sqrt{1-\langle\sigma_i^z\sigma_{i+1}^z\rangle+2 \langle\sigma_i^x\sigma_{i+1}^x\rangle}
	+\sqrt{1-\langle\sigma_i^z\sigma_{i+1}^z\rangle-2 \langle\sigma_i^x\sigma_{i+1}^x\rangle}\right) & \quad -1<\Delta<1,\cr
    1+\langle\sigma_i^z\sigma_{i+1}^z\rangle+\sqrt{(1-\langle\sigma_i^z\sigma_{i+1}^z\rangle)^2-4 \langle\sigma_i^x\sigma_{i+1}^x\rangle^2} & \qquad others.\nonumber
	\end{array}
	\right.
\end{eqnarray}
\end{widetext}

\begin{figure*}[htbp!]
\includegraphics[width=7.2in]{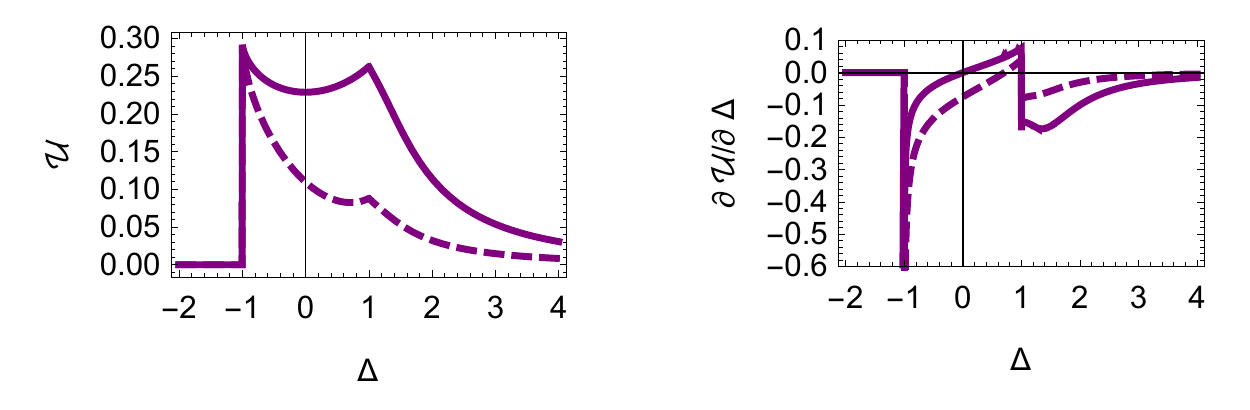}\\
\centering
\caption{(Color online) Local quantum
uncertainty (left) for the
$XXZ$ model: the nearest (solid line) and the next to nearest
neighbor (dashed line) spins and its derivatives (right) vs $\Delta$.}
\label{fig4}
\end{figure*}

Fig. (\ref{fig4}) shows that the local quantum uncertainty
in the region $\Delta>1$ decreases monotonously. There are two
critical points at $\Delta=-1$ and $1$,
which show the phase transitions in both
the nearest and the next to nearest neighbor spin correlations.

\section{LMG model}
We now consider the LMG model \cite{Lipkin1965},
which describes a two-level Fermi system
$\{|+\rangle, |-\rangle \}$, with each
level having degeneracy $g$. The Hamiltonian
for LMG model can be written as
\begin{widetext}
\begin{equation}
	H=\lambda\sum_{m=1}^g\frac12(c_{+m}^\dagger c_{+m}-c_{-m}^\dagger c_{-m})
	-\frac{1}{2N}\sum_{m,n=1}^g
	(c_{+m}^\dagger c_{-m}c_{+n}^\dagger c_{-n}+ c_{-n}^\dagger c_{+n}c_{-m}^\dagger c_{+m}),
\end{equation}
\end{widetext}
where the operators $c_{+m}^\dagger$
and $c_{-m}^\dagger$ create a particle
in the upper and lower levels, respectively.
Alternatively, the LMG model can be
seen as a one-dimensional ring of
spin-1/2 particles with an infinite
range interaction between pairs. In
fact, the Hamiltonian can be rewritten as
\begin{equation}
	H=\lambda S_z -\frac1N(S_x^2-S_y^2),
\end{equation}
with $S_z=\sum_{m=1}^N\frac12(c_{+m}^\dagger
c_{+m}-c_{-m}^\dagger c_{-m})$ and
$S_x+i S_y=\sum_{m=1}^N c_{+m}^\dagger c_{-m}$ \cite{Ring2004}. The LMG model experiences
a second-order quantum phase transitions
at $\lambda=1$.
As $g\rightarrow\infty$, the ground
state, as given by the HF approach,
reads as
\begin{eqnarray}
	|HF\rangle=\prod\limits_{m=1}^\omega a_{0m}^\dagger|-\rangle,
\end{eqnarray}
where the new levels labeled
by $0$ and $1$ are governed by the operators
\begin{eqnarray}\label{am}
	a_{0m}^\dagger&=&\cos\alpha\, c_{-m}^\dagger
	+\sin\alpha\, c_{+m}^\dagger,\nonumber\\
	a_{1m}^\dagger& =&-\sin\alpha\, c_{-m}^\dagger +\cos\alpha\, c_{+m}^\dagger.
\end{eqnarray}
In Eq.(\ref{am}), $\alpha$ is a variational
parameter to be adjusted in order to minimize
energy, which is achieved according to the
choice
\begin{eqnarray}
\begin{split}
	\lambda<1 \Rightarrow \cos2\alpha=\lambda,\\
	\lambda\ge1\Rightarrow \alpha=0.
\end{split}
\end{eqnarray}

Despite being an approximation, the
$HF$ ground state provides the exact
description of the critical point.
The pairwise density operator for
general modes $i=(+m)$ and $j=(-n)$
is described as
\begin{eqnarray}\label{rho}
	\rho_{i,j}=\left(
	\begin{array}{cccc}
		\langle M_i M_j\rangle & 0 & 0 & 0\cr
		0 & \langle M_i N_j\rangle & \langle c_i^\dagger c_j\rangle & 0\cr
		0 & \langle c_j^\dagger c_i\rangle & \langle N_i M_j\rangle & 0\cr
		0 & 0 & 0 & \langle N_i N_j\rangle
	\end{array}
	\right),
\end{eqnarray}
where $M_k=1-N_k$ and $N_k=c_k^\dagger c_k$,
with $k=i,j$. The Eq. (\ref{rho}) shows a
$Z_2$ symmetry.

The evaluated matrix elements of
$\rho$ for the HF ground state are given by
\begin{eqnarray}
	\langle M_{+m} M_{-n}\rangle
	&=& \sin^2\alpha\cos^2\alpha(1-\delta_{mn}),\nonumber\\
	\langle M_{+m} N_{-n}\rangle
	&=& \cos^2\alpha\delta_{mn}+\cos^4\alpha
	(1-\delta_{mn}),\nonumber\\
	\langle N_{+m} M_{-n}\rangle
	&=& \sin^2\alpha\delta_{mn}+\sin^4\alpha
	(1-\delta_{mn}),\nonumber\\
	\langle N_{+m} N_{-n}\rangle
	&=& \sin^2\alpha\cos^2\alpha(1-\delta_{mn}),\nonumber\\
	\langle c_{+m}^\dagger c_{-n}\rangle
	&=& \sin\alpha\cos\alpha\delta_{mn},\nonumber\\
	\langle c_{-n}^\dagger c_{+m}\rangle
	&=& \sin\alpha\cos\alpha\delta_{mn}.
\end{eqnarray}
Here for $m\ne n$, the density matrix
Eq. (\ref{rho}) is diagonal and
the state is completely pairwise
uncorrelated. On the other hand, for
$m=n$, there are quantum correlations
between the modes. These correlations
vanish for $\lambda>1$, which gives rise to the fully
polarized states.

i) Negativity of LMG. We can derive the analytical expression
of negativity for the LMG model
\begin{equation}
	\mathcal{N}=\frac12\sqrt{1-\lambda^2}.
\end{equation}

ii) Information deficit for LMG.
The analytical expression of information
deficit for the LMG model has the following form,
\begin{equation}
	\mathcal{I}=-\frac{\log \left(\frac{1-\lambda }{4}\right)+\log (\lambda +1)}{2}- \lambda  \tanh ^{-1}(\lambda )/\ln2.
\end{equation}
Here, the optimal measurement is arrived at
$\theta=\phi=0$ for Eq. (\ref{bases}).


iii) By tedious calculation, we have
the expressions of trace distance discord
for the LMG model,
\begin{equation}
	\mathcal{D}=\frac12\sqrt{1-\lambda^2},
\end{equation}
which is the same as the measures negativity.

iv) Local quantum uncertainty for LMG.
By straightforward computation, we obtain the analytical expressions
of the local quantum uncertainty for LMG,
\begin{equation}
	\mathcal{U}=1-\lambda.
\end{equation}

\begin{figure*}[htbp!]
\includegraphics[width=6.9in]{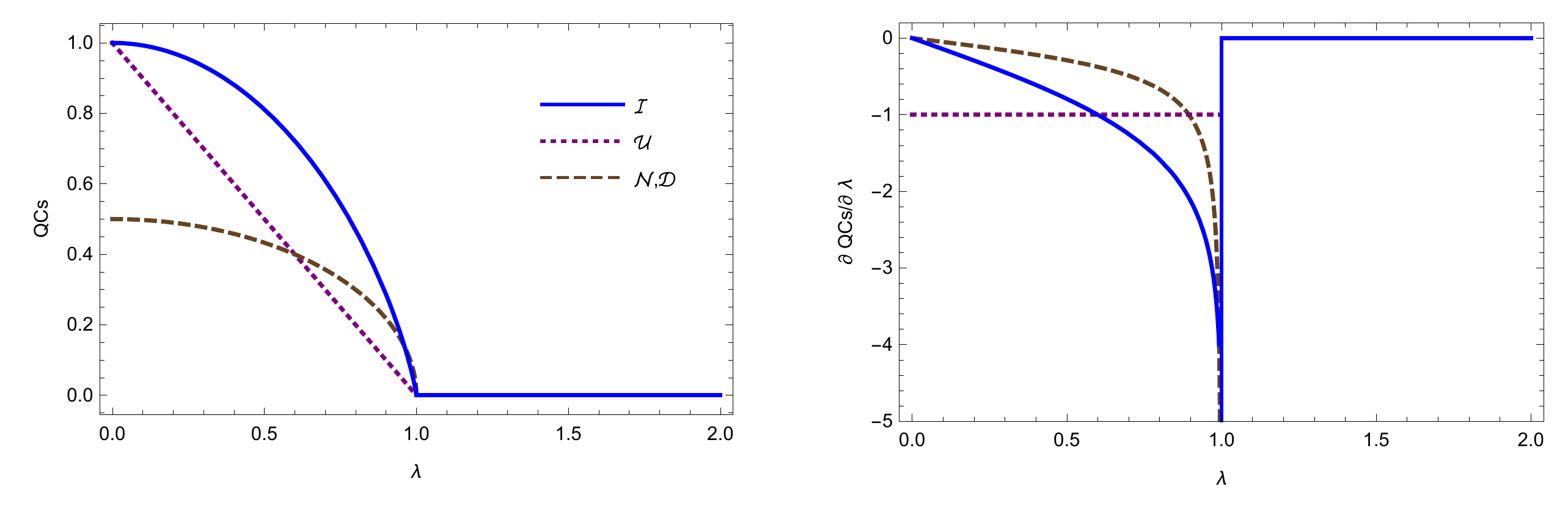}\\
\centering
\caption{(Color online) All kinds quantum
correlations (QCs) and their derivative
vs parameter $\lambda$ between
modes $(+m)$ and $(-m)$ in the HF ground state of the LMG
model. }
\label{fig5}
\end{figure*}

From the above analytical results,
we see that all of the four kinds of quantum correlations decrease monotonously with
$\lambda<1$, see Fig.(\ref{fig5}). And in the region
$\lambda\ge1$ the quantum correlations vanishes.
One can also observe that the derivatives
of the four kind quantum correlations
exhibit a signature of the quantum phase
transition (see the right figure in Fig. (\ref{fig5}).

\section{Conclusions}
We have investigated the behavior of quantum
correlations for the Heisenberg $XXZ$ spin-1/2 chain via negativity, information
deficit, trace distance discord and local quantum uncertainty.
Some properties of the $XXZ$ system are given as follows:

(1) Information deficit and local quantum uncertainty demonstrate
the quantum phase transitions at $\Delta=-1$ and $1$.
However, the negativity and trace distance discord fail to
detect the quantum phase transition at $\Delta=1$.

(2) For the nearest neighbor spins, entanglement and information deficit reach their
maximal value at $\Delta=1$.
While trace distance discord has the maximal value at $\Delta=0$, and
local quantum uncertainty has the maximal value at $\Delta=-1$.

(3) The quantum correlation of the nearest neighbor spins are
greater than that of the next to nearest neighbor spins for all the
four kinds measures.

We have also considered quantum correlations in the Hartree-Fock ground state
of the Lipkin-Meshkov-Glick model.
All the four quantum correlation measures has been analytically
worked out. The behaviors for both $XXZ$ spin-1/2 chains and LMG model
have been discussed. It is shown that all of the quantum correlation measures
exhibit signatures of the quantum phase transitions.

We have studied the ability of quantum correlations
to spotlight critical points of quantum phase
transitions for an infinite spin chain described by the
$XXZ$ model, in terms of four distinct types of
quantum correlations between pairs of
nearest and next to nearest neighbor
spins. All the measures of these quantum correlations
show quantum phase transitions
at $\Delta=-1$ for both the nearest
and next to nearest neighbor spins.
However, the information deficit and local quantum uncertainty
exhibit one more singularity at the critical point $\Delta=1$.
It can be seen that information deficit
and local quantum uncertainty are better
in studying the critical points for the $XXZ$ spin system.

Our work highlights the type of quantum
correlations involved at the quantum phase transition of the
$XXZ$ system and LMG model. These measures show the ways to explore quantum phase
transition in condensed matter physics.
These quantities are important tools
in the investigation of quantum phase transitions
in realistic experimental scenarios.
The approach can be also used to explore quantum
phase transitions in other physical systems.

\section*{Acknowledgments}
This work is supported by the NSFC (11675113, 11765016)
and Jiangxi Education Department Fund (GJJ161056, KJLD14088).


\end{document}